\documentclass[a4paper]{article}
\usepackage{fullpage,geometry}
\usepackage{graphicx} 
\usepackage{siunitx}
\usepackage[square,comma]{natbib}
\usepackage{amsmath,amssymb}
\usepackage[english]{babel}
\usepackage{todonotes}
\usepackage[unicode, pdftex,breaklinks=true]{hyperref}
\title{Separating the spectral counterparts in NGC 1275/Perseus cluster in X-rays}
\date{}
\author{Elena Fedorova $^{1,2}$, Lidiia Zadorozhna $^{3,4,5}$, Anatolii Tugay $^{3}$, \\ Nadiia Pulatova $^{6,7}$, Alexander Ganz $^{4}$, Olexandr Gugnin $^{2}$}

\begin{document}

\maketitle

$^{1}$ \quad INAF - Osservatorio Astronomico di Roma, Frascati str. 33, Monte Porzio Catone, 00040, Rome, Italy;\\
$^{2}$ \quad Astronomical Observatory Taras Shevchenko National University of Kyiv, Observatorna str. 3, 04053, Kyiv, Ukraine;\\
$^{3}$ \quad Taras Shevchenko National University of Kyiv, Hlushkova ave. 4, 03127, Kyiv, Ukraine;\\
$^{4}$ \quad Jagiellonian University,
Faculty of Physics, Astronomy and Applied Computer Science, ul. prof. Stanis{{\l}}awa {{\L}}ojasiewicza 11, 30-348, Krak{{\'o}}w, Poland;\\
$^{5}$ \quad Niels Bohr Institute, Jagtvej 155A, 2200, K{{\o}}benhavn, Denmark;\\
$^{6}$ \quad Main Astronomical Observatory of the National Academy of Sciences of Ukraine, Akademika Zabolotnoho str. 27, 03143, Kyiv, Ukraine;\\
$^{7}$ \quad  Max-Planck-Institut f{\"u}r Astronomie, K{\"o}nigstuhl 17, D-69117, Heidelberg, Germany. 

\bigskip
Correspondence: Elena Fedorova elena\_f@mail.univ.kiev.ua;\\
Lidiia Zadorozhna lidiia.zadorozhna@knu.ua, zadorozhna@nbi.ku.dk

\begin{abstract}
    We develop the recipe to separate the spectral counterparts of the AGN NGC 1275 from the emission of the Perseus cluster surrounding it in the spectra observed by Suzaku/XIS cameras with no usage of the spectral fitting models. The Perseus cluster emission reaches higher energies than is typical for the most AGN-situated dense surroundings (i.e. up to 9-10 keV). That is why the separation between the AGN and cluster spectra is especially important in this case. To avoid the degeneracy due to the huge quantity of the spectral fitting parameters such as abundances of elements the cluster consists of, thermal and Compton emission of the nucleus itself, and the jet SSC/IC emission spectral parameters as well we prefer to avoid the spectral fitting usage to perform this task. Instead, we use the spatial resolution of the components and double background subtracting. For this purpose we choose the following regions to collect all the photons from them: (1) circular or square-shaped region around the source (AGN); (2) ring-shaped (or non-overlapped square) region surrounding the AGN (for cluster); (3) remote empty circular region for the background. Having collected the photons from those regions we subtract the background (i.e. photons from the third region) from the source and cluster spectra. Next, we subtract the re-normalized cluster counts from the AGN spectrum; using the relation between the emission line amplitudes in the AGN and cluster spectra as the renormalization coefficient. 
We have performed this procedure on the whole set of the Suzaku/XIS observational data for NGC 1275 to obtain the cleaned spectra and light curve of the AGN emission in this system.
\end{abstract}

\section{Introduction}
\label{sec1}

NGC 1275 (3C 84 or Perseus A) is a giant elliptical galaxy hosting the active nucleus of the Seyfert 1.5 type \citep{Veron2010} classified as a Fanaroff-Riley type I radio-loud (RL) with a compact central source and extended jet \citep{Buttiglione2010, Vermeulen1994}. It is situated at the center of the Perseus/Caldwell24 cluster of galaxies with the redshift z=0.01756 \citep{1999PASP..111..438F}. The supermassive black hole (SMBH) mass in the AGN center is of $\sim 3.4 \cdot 10^8~M_\odot$ \citep{Wilman2005}. 

 NGC 1275 is surrounded by a massive network of cool gaseous filaments with a frozen-in magnetic field associated with a cooling flow \citep{1970ApJ...159L.151L,2018AAS...23125211C}. These filaments emit an intensive network of spectral lines. Moreover, NGC 1275 contains the high-velocity system (HVS) of molecular gas moving in front of it towards its center \citep{1974ApJ...189L..59F,2011MNRAS.418.2154F}, and the hot radio-bright bubbles of relativistic plasma fed by the jets, heating the inner gas and preventing the radiative cooling-induced runaway \citep{2008A&A...484..317S}. The HVS system probably plays a significant role in the nucleus fueling \citep{2008ApJ...672..252L}.

The first detailed view of NGC 1275 and the Perseus Cluster in X-rays was performed by RoSat (Roentgen Satellite)/HRI (High-Resolution Imager) \citep{1993MNRAS.264L..25B}. It revealed the complicated substructure of the X-ray surface brightness within $\approx$ 5 arcminutes around the NGC 1275 center. They also found the 30$\%$ surface brightness variations presumably connected with the inflating of the relativistic plasma bubbles by the jets. Next, in 1996 NGC 1275 was detected by Atmospheric Cherenkov Telescopic System SHALON at TeV energies \citep{2013JPhCS.409a2111S}. 

The SMBH nested in the nucleus of NGC 1275 was the first black hole "heard" by the Chandra X-ray Observatory \citep{2003cxo..pres...17.}. These observations had detected the sound waves expanding in the warm gas surrounding the nucleus and heating it; they also revealed two vast bubbles in this gas, emanating from the NGC 1275 "central engine" \citep{2003MNRAS.344L..43F}.

Since then NGC 1275 has been observed in X-rays quite intensively, by several cosmic X-ray missions such as XMM-Newton, Suzaku, Swift, etc. Mutual analysis of the Suzaku/XIS (X-ray Imaging Spectrometer) and Fermi/LAT (Large Area Telescope) data detected the first evidence of correlated variability in the X-rays and $\gamma$-rays during the period 2013-2015. Wide-band variability of NGC 1275 was investigated in \citep{2017ApJ...848..111B,2010ApJ...715..554K} ($\gamma$-rays),  \citep{galaxies8030063} (X-rays to $\gamma$-rays), \citep{2021ApJ...906...30I} (UV no soft $\gamma$-rays), \citep{2014A&A...564A...5A} (radio to the highest $\gamma$-rays). Short-term, rapid fluctuations in UV to X-ray emissions show a correlation with GeV $\gamma$-rays. The gradual correlated changes of the UV and soft/hard X-ray fluxes along with the GeV $\gamma$-rays were confirmed \citep{2021ApJ...906...30I}.

The $\gamma$-ray activity in 3C 84 can be split into short-term outbursts on top of a long-term, slowly increasing trend (also seen in the optical V band by Nesterov et al. \citep{1995A&A...296..628N}).  

Long-term variations have been detected in the soft X-ray range, alongside variations in radio and GeV $\gamma$-ray emissions (as noted in Fabian et al. \citep{2015MNRAS.451.3061F} and Fukazawa et al. \citep{2018ApJ...855...93F}). These fluctuations likely suggest that we are witnessing the presence of X-ray emissions originating from the jet \citep{2021ApJ...906...30I}.

At the highest energies (90-1200 GeV, Major Atmospheric Gamma Imaging Cherenkov Telescopes (MAGIC)) the spectrum follows the power-law with the photon index above 3 attributed to pure jet with no additional components due to diffuse medium \citep{2016A&A...589A..33A}; the flares were observed in this energy range as well \citep{2018A&A...617A..91M}. Such flares were interpreted in \citep{2019Galax...7...72B} as the result of the jet precession. 

The best spectral resolution was reached in the Hitomi observation of NGC 1275 \citep{2018PASJ...70...13H}; despite the presence of the extremely bright emission line near 6.6 keV, the narrow F-K$\alpha$ 6.4 keV line with the EW$\approx$20-25 eV emitted either in the outer parts of the accretion disk or in the torus was resolved as well. However the iron K$\alpha$ widths detected earlier by XMM-Newton was significantly higher: $\approx$ 70-80 eV in 2006 \citep{2013PASJ...65...30Y} and even $\approx$165 eV in 2001 \citep{2003ApJ...590..225C}. The recent Swift data analyses \citep{2021ApJ...906...30I} confirmed the presence of the Fe-K fluorescence line indicating rather the presence of an accretion disk than the torus, but the low quality of the BAT (15-150 keV) spectrum not allowed them to detect high-energy exponential cut-off in the hard X-ray spectrum. 

Absorption features were identified in the radio continuum emitted by the central parsec-scale jet \citep{Nagai2019}. Their blueshift in the range of 300-600 km/s indicates a rapid molecular outflow originating from the AGN with the column density $N_{H_{2}}\approx 2.3\cdot 10^{22}~{\rm cm^{-2}}$.

The deep Cycle-19 of the Chandra/HETG (High Energy Transmission Grating) observations of NGC 1275 had revealed several emission lines of highly ionized iron, mildly ionized Mg, S, and Si \citep{2021arXiv210804276R}). The 6.4 keV iron-K$_{\alpha}$ line was detected by Chandra as well, with even lower EW at the 14 eV level. 
   
Multiwavelength study of NGC 1275 in \citep{2021MNRAS.503..446G} revealed the four distinct activity phases in X-ray and $\gamma$ with an increase of the baseline X-ray and UV fluxes during the first three phases, explained by the inverse Comptonization of the synchrotron photons by the jet electrons \citep{2021MNRAS.503..446G}. 

Due to the presence of the near year-long quasi-periodic oscillations at 1.3 mm wavelength, NGC 1275 is also considered as a possible double BH candidate \citep{2022ApJ...925..207Z}.

The spectrum of NGC 1275 is contaminated extremely with the Inter-Cluster Medium (ICM) emission contributing up to 80-90\% to it within the soft to intermediate X-rays. In this work, we develop an algorithm of two-step background subtracting which could give us the possibility to subtract the cluster emission from the AGN spectrum with no usage of the spectral modeling.
In the section {\bf \ref{sec2} Data processing} we describe the Suzaku/XIS data reduction and the recipe we propose to perform two-step background subtracting. In section {\bf \ref{sec3} Fitting the spectra} we demonstrate the resulting cleaned spectra of the nucleus and perform the simple preliminary fitting of them using the power-law model with the neutral absorption \textit{po*tbabs}. The resulting monthly light curve covering the timespan of nearly 10 years is also shown in this section. In the section {\bf \ref{sec4} Conclusions and further perspectives} we describe some further steps we are going to pass using these results to obtain the physically motivated picture of the NGC 1275 nucleus and its high-energy spectrum and draw out our conclusions.

\section{Data processing}
\label{sec2}

The main goal of this work is to develop a model-independent recipe for separating the spectra of the source from the spectra of its surroundings with no means of spectral fitting. This task was performed in two steps, the first one represents the standard data reduction to obtain the uncleaned spectra of the source (in our case the AGN + cluster), its surroundings (i.e. the cluster), and the remote background. The first (standard reduction) step is described in the Subsection \ref{sec2.1}. In the second step, we describe the recipe and apply it to the NGC 1275 X-ray spectra to separate the AGN counterparts from those of the surrounding cluster. The second step of our data treatment is described in Subsection \ref{sec2.2}. 

\subsection{Primary data reduction}
\label{sec2.1}

\begin{figure*}[t]
\centering
\includegraphics[width=.45\textwidth]{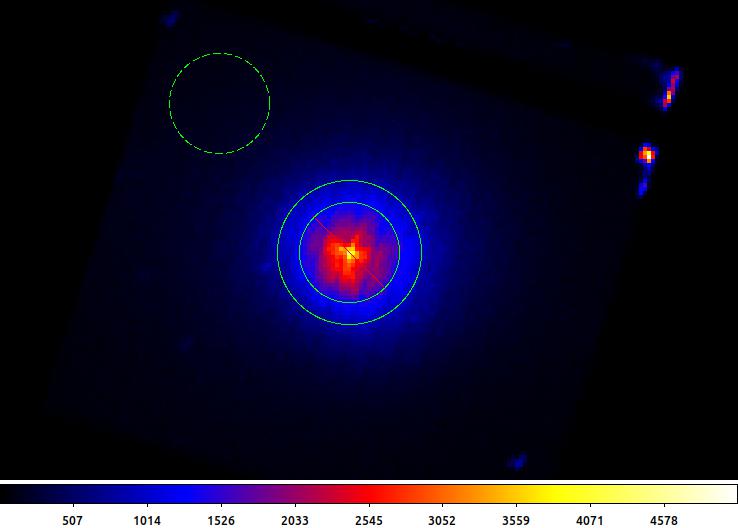}
\includegraphics[width=.45\textwidth]{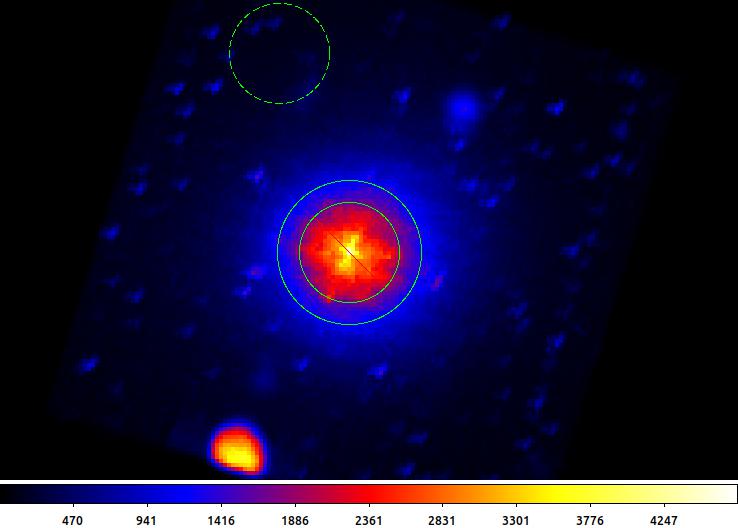}
     \caption{The 0.1-10 keV images of NGC 1275 by Suzaku/XIS for the observational window 101012010 (August 2006) with the regions from where the source, cluster, and background counts were extracted. Left: the image from the XIS0 camera, AGN circle, cluster ring, and the remote background region (green circle); right: the XIS1 camera image, the AGN circle, cluster, and remote background region.} 
\label{imsirc}  
\end{figure*}

The images and spectra were obtained using the standard multi-mission photon-collecting procedure {\it xselect}. After the image creation, the following regions were selected and stored in the *.reg files using the SAOImageDS9 astronomical imaging and data visualization application by the Chandra X-ray Science Center (CXC), the High Energy Astrophysics Science Archive Center (HEASARC) and the JWST Mission office at Space Telescope Science Institute \citep{2003ASPC..295..489J}: for the source (circular region with AGN in the center), for the surrounding cluster (ring region centered on the AGN), and for the remote background (empty circular region as far away as possible from the AGN). Taking into account that the PSF half-brightness radius is near 1 arcmin \citep{2013A&A...552A..47K} and thus the recommended radius of the source extraction region is not smaller than 1.5 arcmin, we choose the following areas of the extraction: 1.74 arcmin-radii circles for both the AGN and outer backgrounds (i.e. 100 ds9 physical units) and the areas with the inner radius of 1.74 arcmin (100 ds9 physical units) and the outer radii of 2.57 arcmin (i.e. 144 ds9 physical units). The examples of the images and regions are shown in Fig.\ref{imsirc}.

\begin{figure*}[t]
\centering
\includegraphics[width=.45\textwidth]{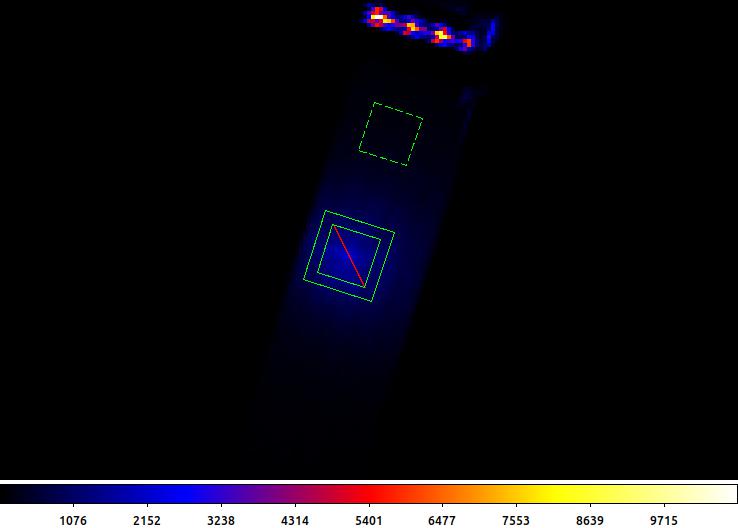}
\includegraphics[width=.45\textwidth]{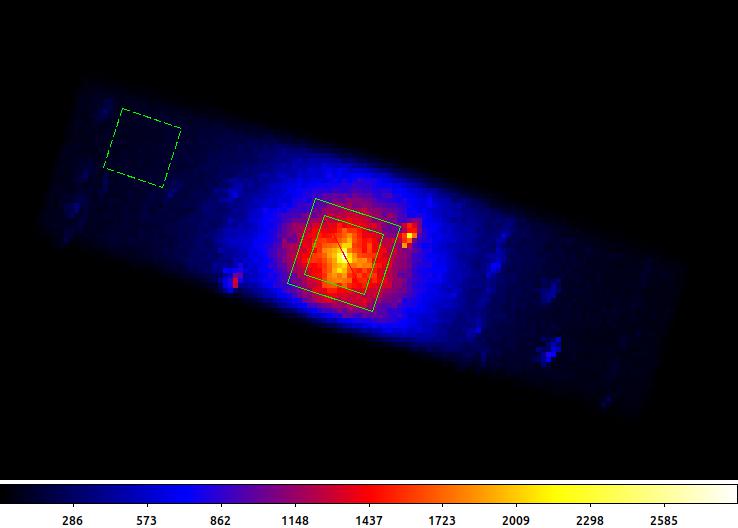}
     \caption{The 0.1-10 keV images of NGC 1275 by Suzaku/XIS for the observational window 107006010 (August 2012) with the regions from where the source, cluster, and background counts were extracted. Left: XIS0 camera, right: XIS1 camera.} 
\label{imsquare}  
\end{figure*}

Some observations were performed in partial window mode. For such ones, we had chosen the square-shaped regions with 2 arcmin size for the AGN and the remote background (empty square region as far away as possible from the AGN), and 2.88 arcmin size for the outer square region with inner AGN-centered region excluded for the cluster. The examples of the images and regions are displayed in Fig.\ref{imsquare}. 

\begin{figure}[!htb]
\centering
\includegraphics[width=1.0 \textwidth]{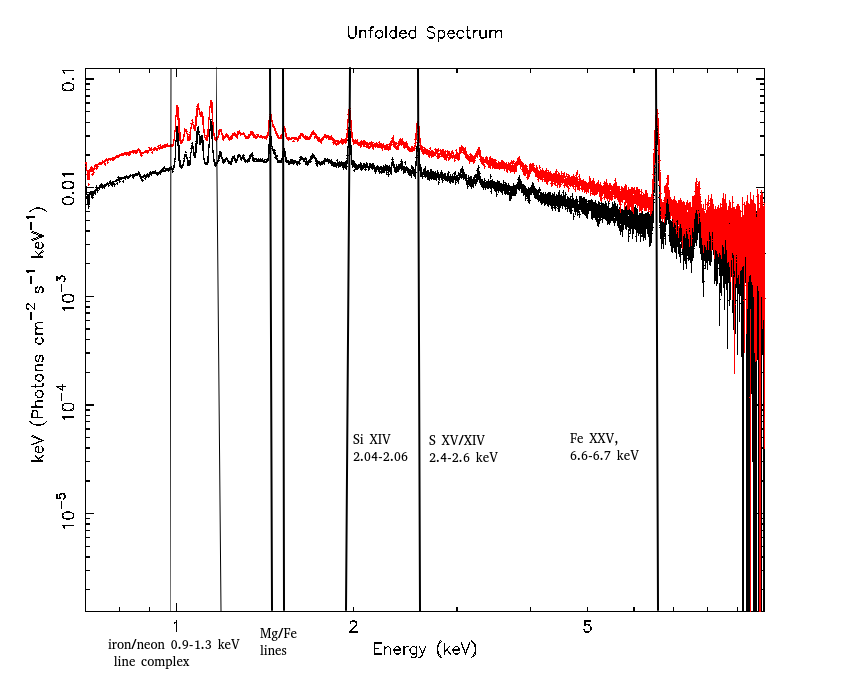}
     \caption{The example of the 0.5-10.0 keV spectra of the surroundings (cluster) and NGC 1275 (uncleaned AGN + cluster) extracted in the standard way for the observational window 101012010. The uncleaned AGN + cluster spectrum extracted from the circular central region around the AGN is shown in red, and the surrounding cluster spectrum extracted from the ring-shaped region is shown in black.} 
\label{spectrum}  
\end{figure}

After processing the images and creating the regions, we extracted the spectra of the source, cluster, and background for all the available cameras. The spectra of the same object obtained for different cameras were merged into one single spectrum per observation. The resulting unified XIS spectra were obtained by merging the spectra for every camera into one using the {\it addascaspec} routine included in the HEASoft X-ray data analysis and processing off-line software package together with the corresponding response matrices and ancillary files. The example of the resulting spectra for the AGN circle region and the surrounding cluster (a ring region centered on the AGN) is depicted in Fig. \ref{spectrum}.  

We also merged the spectra of the closed-in-time observations (i.e. within one month) in the same way and obtained the resulting 20 files for different observational periods. The observations that were merged are shown in Table \ref{xislog} together in one cell. 

\subsection{Distinguishing cluster and AGN components: the routine}
\label{sec2.2}

After the first data reduction step we obtained the source (uncleaned AGN + cluster spectra), surrounding cluster, and background for every period of observations. In the source and cluster spectra, the sets of emission lines typical for collisionally-ionized diffuse gas are clearly visible. Following the data and emission models \textit{bapec} provided by atomdb (Atomic Data for Astrophysicists, \url{http://www.atomdb.org/}) calculated by means of the HULLAC code by \citep{1995ApJ...438L.115L} these lines can be interpreted as follows:\begin{itemize}
    \item{the set of iron and/or neon lines within the energy range 0.9-1.3 keV (for more details see the Table \ref{apeclines0});}
    \item{two lines of magnesium and iron near 1.5 keV (for more details see the Table \ref{apeclines1});}
    \item{Si XIV line at 2.0 keV (for more details see the Table \ref{apeclines2});}
    \item{S XV/XIV line near 2.5 keV (for more details see the Table \ref{apeclines3});}
    \item{Fe XXV lines near 6.7 keV (for more details see Table \ref{apeclines4}).}
\end{itemize}

We can use these lines and relations between their amplitudes as clues to separate the spectral counterpart of the cluster from that of the active nucleus. Namely, we can choose several single lines easily distinguishable from others (in our case these are: Si XIV near 2 keV, S XV/XIV at 2.4-2.6 keV, and Fe XXV near 6.7 keV) and calculate the relations between their amplitudes for the source and the cluster regions. The equivalence of these relations within the error bars tells us that the lines visible in the source and cluster areas spectra are of the same origin and thus are emitted by the same medium. Otherwise, if some relation is higher for a source region this can be considered as a sign of the presence of the line emission in the AGN spectrum too (more often this concerns the case of the Fe XXV line as near it we can often observe the Fe-K$\alpha$ 6.4 keV line emitted from the AGN "central engine"). Additionally, from the results shown in Table\ref{linerel}, we can see that the relations between A3 and A1 or A2 are compatible with each other for AGN + cluster and cluster ring regions. Therefore we can conclude that the emission in Fe-K lines is not significant here, and thus the spectrum of AGN in NGC 1275 was rather jet-dominated in the course of the SUZAKU/XIS observations.  

\begin{center}
\begin{table}
\centering
    \caption{Relations between amplitudes of the emission lines in the cluster spectra. A1 is the amplitude of Si XIV 2 keV line, A2 is that of S XV/XVI within 2.4-2.6 keV and A3 is that of Fe XXV line near 6.7 keV.}
\centering
  \begin{tabular}{|p{40pt}|p{50pt}|p{50pt}|p{50pt}|p{50pt}|p{50pt}|p{50pt}|} \hline
        obs. & \multicolumn{3}{|c|}{AGN + cluster}  & \multicolumn{3}{|c|}{cluster} \\\cline{2-7} 
        date & A$_1$/A$_2$ & A$_1$/A$_3$, & A$_3$/A$_2$ & A$_1$/A$_2$ & A$_1$/A$_3$, & A$_3$/A$_2$ \\\hline
        2006-02 & 2.11$\pm$0.2 & 1.38$\pm$0.12 & 1.53$\pm$0.15 & 2.07$\pm$0.21 & 1.27$\pm$0.14 & 1.62$\pm$0.17 \\\hline 
        2006-08 & 2.0$\pm$0.2 & 1.48$\pm$0.08 & 1.36$\pm$0.05 & 1.9$\pm$0.2 & 1.45$\pm$0.08 & 1.31$\pm$0.05 \\\hline
        2007-02 & 2.3$\pm$0.25 & 1.66$\pm$0.2 & 1.4$\pm$0.1 & 2.25$\pm$0.27 & 1.56$\pm$0.21 & 1.44$\pm$0.11 \\\hline
        2007-08 & 2.0$\pm$0.2 & 1.11$\pm$0.13 & 1.81$\pm$0.19 & 2.09$\pm$0.24 & 1.16$\pm$0.14 & 1.80$\pm$0.20 \\\hline
        2008-02 & 2.0$\pm$0.1 & 1.48$\pm$0.05 & 1.36$\pm$0.04 & 1.9$\pm$0.13 & 1.45$\pm$0.06 & 1.31$\pm$0.06 \\\hline
        2008-08 & 1.5$\pm$0.25 & 1.24$\pm$0.16 & 1.21$\pm$0.15 & 1.54$\pm$0.27 & 1.13$\pm$0.17 & 1.36$\pm$0.19 \\\hline
        2009-02 & 1.64$\pm$0.23 & 1.33$\pm$0.15 & 1.23$\pm$0.14 & 1.57$\pm$0.3 & 1.34$\pm$0.16 & 1.17$\pm$0.15 \\\hline
        2009-08 & 2.09$\pm$0.2 & 1.25$\pm$0.1 & 1.67$\pm$0.15 & 2.12$\pm$0.2 & 1.28$\pm$0.1 & 1.65$\pm$0.15 \\\hline
        2010-02 & 1.83$\pm$0.28 & 1.33$\pm$0.20 & 1.38$\pm$0.21 & 1.64$\pm$0.29 & 1.28$\pm$0.2 & 1.28$\pm$0.21 \\\hline 
        2010-08 & 1.66$\pm$0.25 & 1.21$\pm$0.16 & 1.35$\pm$0.16 & 1.79$\pm$0.26 & 1.38$\pm$0.19 & 1.32$\pm$0.16\\\hline 
        2011-02 & 2.06$\pm$0.15 & 1.23$\pm$0.09 & 1.68$\pm$0.10 & 2.19$\pm$0.16 & 1.31$\pm$0.10 & 1.68$\pm$0.11 \\\hline
        2011-07 & 1.47$\pm$0.17 & 1.19$\pm$0.14 & 1.24$\pm$0.15 & 1.62$\pm$0.2 & 1.21$\pm$0.15 & 1.33$\pm$0.19 \\\hline
        2011-08 & 2.0$\pm$0.3 & 1.3$\pm$0.3 & 1.5$\pm$0.3 & 1.8$\pm$0.3 & 1.1$\pm$0.3 & 1.7$\pm$0.3 \\\hline
        2012-02 & 2.13$\pm$0.25 & 1.57$\pm$0.23 & 1.36$\pm$0.21 & 2.18$\pm$0.31 & 1.62$\pm$0.24 & 1.35$\pm$0.22\\\hline 
        2012-08 & 1.83$\pm$0.24 & 1.26$\pm$0.16 & 1.27$\pm$0.15 & 1.77$\pm$0.25 & 1.2$\pm$0.17 & 1.28$\pm$0.17 \\\hline
        2013-02 & 2.26$\pm$0.18 & 1.41$\pm$0.15 & 1.61$\pm$0.16 & 2.25$\pm$0.2 & 1.33$\pm$0.16 & 1.69$\pm$0.18\\\hline
        2013-08 & 1.95$\pm$0.23 & 1.45$\pm$0.19 & 1.35$\pm$0.18 & 1.89$\pm$0.28 & 1.60$\pm$0.20 & 1.20$\pm$0.19 \\\hline
        2014-02 & 1.94$\pm$0.23 & 1.35$\pm$0.22 & 1.44$\pm$0.22 & 1.85$\pm$0.3 & 1.22$\pm$0.25 & 1.5$\pm$0.2 \\\hline
        2014-08 & 1.9$\pm$0.34 & 1.55$\pm$0.3 & 1.23$\pm$0.28 & 1.84$\pm$0.42 & 1.1$\pm$0.29 & 1.67$\pm$0.37 \\\hline
        2015-03 & 1.75$\pm$0.23  & 1.17$\pm$0.10 & 1.2$\pm$0.11 & 1.62$\pm$0.25 & 1.27$\pm$ 0.12 & 1.28$\pm$0.13 \\\hline
    \end{tabular}
    \vspace{0.05in} \label{linerel}
\end{table}
\end{center}

Using the mean values of the relations between the amplitudes of these lines emitted from the source and cluster areas, we can pass now to the next step of cleaning the source (AGN + cluster) spectra from the contamination by the surrounding cluster counterpart. The presence of bright emission lines in the cluster spectra enables us to perform this with no usage of the spectral fitting models. We can use the background-subtracted cluster spectrum as a correction file for the source spectrum with the correction coefficient set to the mean value of the three relations between emission line amplitudes for source and cluster areas. Together with the correction file, we add the systematic errors corresponding to those of the cluster spectrum with the factor of the correlation coefficient. A more detailed description of the recipe is shown in Appendix \ref{sec7}. The resulting cleaned spectra of the AGN are shown in Fig.\ref{unfspecs}. 

\section{Fitting the spectra}
\label{sec3}

\begin{figure*}[!htb]
\centering
\includegraphics[width=.45\textwidth]{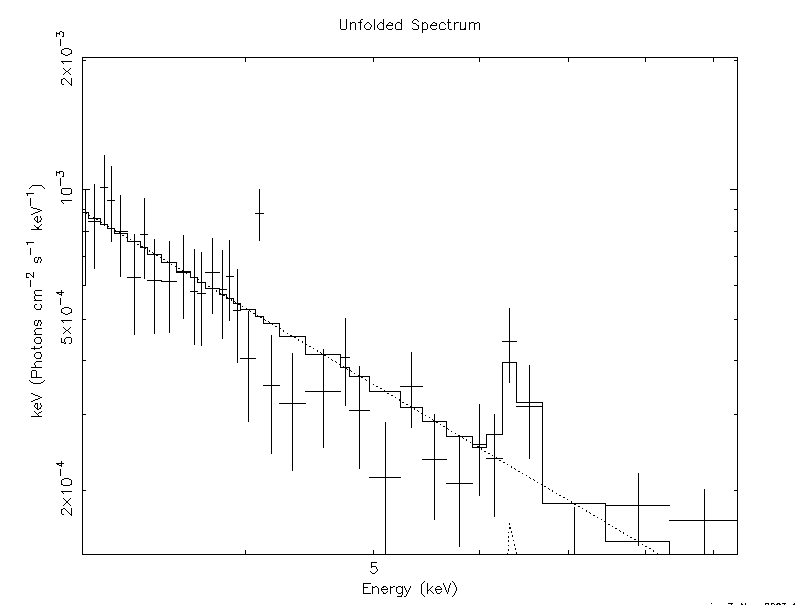}
\includegraphics[width=.45\textwidth]{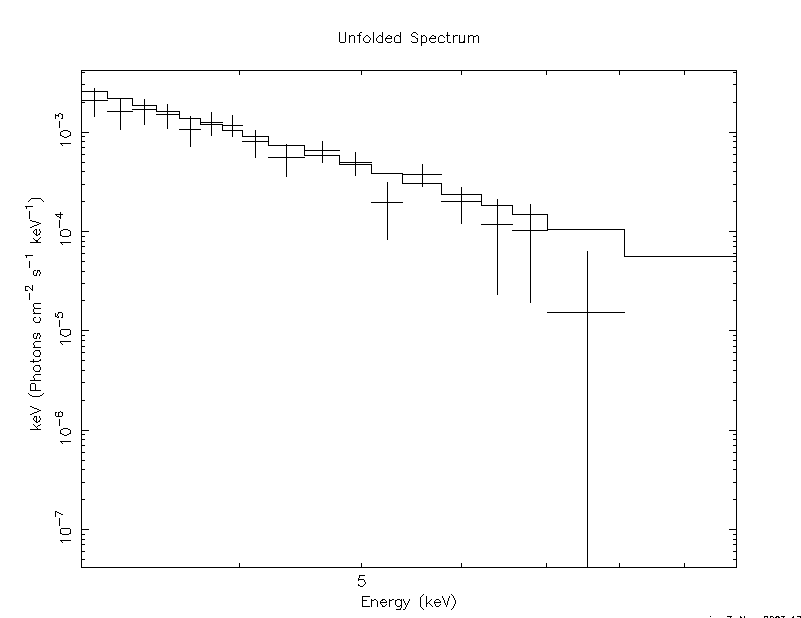}
\caption{The 1-10 keV range unfolded spectra of NGC 1275 by Suzaku/XIS for the observation performed on August 2006 (left) and August 2007 (right) with the models.} 
\label{unfspecs}  
\end{figure*}

Taking into account the probable jet dominance in NGC 1275 in this consideration we confine ourselves with just the simplest spectral model (namely, neutrally absorbed power-law): $po(\Gamma)*tbabs(N_H)$; the spectral fitting was performed using the Xspec v.12.10.1f package of the HEASoft software, version 6.26. 
The results of the fitting are shown in Table\ref{fits}; the two examples of the unfolded spectra with the model are shown also in Fig.\ref{unfspecs}. 

\begin{center}
\begin{table}
\centering
    \caption{Spectral parameters to the Suzaku/XIS cleaned AGN spectra in NGC 1275. The data events from various observations shown in one cell, and more than one observation during one month were merged to produce one common spectrum.}
    \begin{tabular}{|p{50pt}|p{50pt}|p{80pt}|p{40pt}|p{80pt}|} \hline
        obs. & $\Gamma$ & N$_H$, & $\chi^2/d.o.f$ & F$_{2-10}$ \\
        date & & 10$^{22}$ cm$^{-2}$ &  & 10$^{-11}$ erg/cm$^2$sec \\\hline
        2006-02 & 2.8$\pm$0.2 & $<$0.82 & 1280.2/702 & 18.2$\pm$0.2 \\\hline  
        2006-08 & 2.6$\pm$0.2 & 0.10$\pm$0.06 & 442.0/184 & 5.9$\pm$0.4 \\\hline
        2007-02 & 2.6$\pm$0.7 & $<$6.6 & 71.4/121 & 6.3$\pm$1.2\\\hline
        2007-08 & 2.0$_{-0.2}^{+0.3}$ & $<$1.4 &  154.6/68 & 6.6$\pm$0.2\\\hline
        2008-02 & 2.6$\pm$0.1 & 0.4$\pm$0.1 & 154.8/59 & 8.1$\pm$0.3\\\hline
        2008-08 & 2.8$\pm$0.4 & $<$1.6 & 101.8/96 & 5.7$\pm$0.2\\\hline
        2009-02 & 2.4$\pm$0.1 & 0.3$\pm$0.1 & 420/250 & 5.4$\pm$0.4\\\hline
        2009-08 & 2.9$\pm$0.3 & 0.3$\pm$0.1 & 309/157 & 6.0$\pm$0.3 \\\hline
        2010-02 & 2.4$\pm$0.1 & 0.17$\pm$0.06 & 277.5/96 & 7.8$\pm$0.2\\\hline  
        2010-08 & 2.4$\pm$0.1 & 0.14$\pm$0.06 & 284.1/103 & 8.3$\pm$0.2 \\\hline 
        2011-02 & 2.31$\pm$0.05 & 0.29$\pm$0.07 & 300.5/154 & 8.8$\pm$0.2 \\\hline
        2011-07 & 2.18$\pm$0.07 & 0.09$\pm$0.04 & 237.5/97 & 7.3$\pm$0.3 \\\hline
        2011-08 & 2.20$\pm$0.04 & 0.20$\pm$0.03 & 211.0/82 & 14.2$\pm$0.2 \\\hline
        2012-02 & 2.19$\pm$0.05 & 0.10$\pm$0.03 & 264.8/73 & 12.8$\pm$0.2 \\\hline 
        2012-08 & 2.13$\pm$0.04 & 0.18$\pm$0.02 & 263.0/155 & 19.2$\pm$0.2\\\hline
        2013-02 & 2.15$\pm$0.04 & 0.27$\pm$0.15 & 382.8/145 & 14.2$\pm$0.2\\\hline 
        2013-08 & 2.17$\pm$0.03 & 0.25$\pm$0.03 & 448.4/152 & 17.8$\pm$0.2 \\\hline 
        2014-02 & 2.16$\pm$0.03 & 0.13$\pm$0.02 & 274.4/142 & 20.2$\pm$0.3 \\\hline 
        2014-08 & 2.05$\pm$0.04 & 0.20$\pm$0.03 & 246.2/77 & 22.5$\pm$0.2 \\\hline 
        2015-03 & 1.96$\pm$0.03  & 0.25$\pm$0.10 & 365.5/135 & 17.1$\pm$0.2\\\hline
    \end{tabular}
    \vspace{0.05in} \label{fits}
\end{table}
\end{center}

\begin{figure*}[!htb]
   \centering
   \includegraphics[width=1.0\hsize]{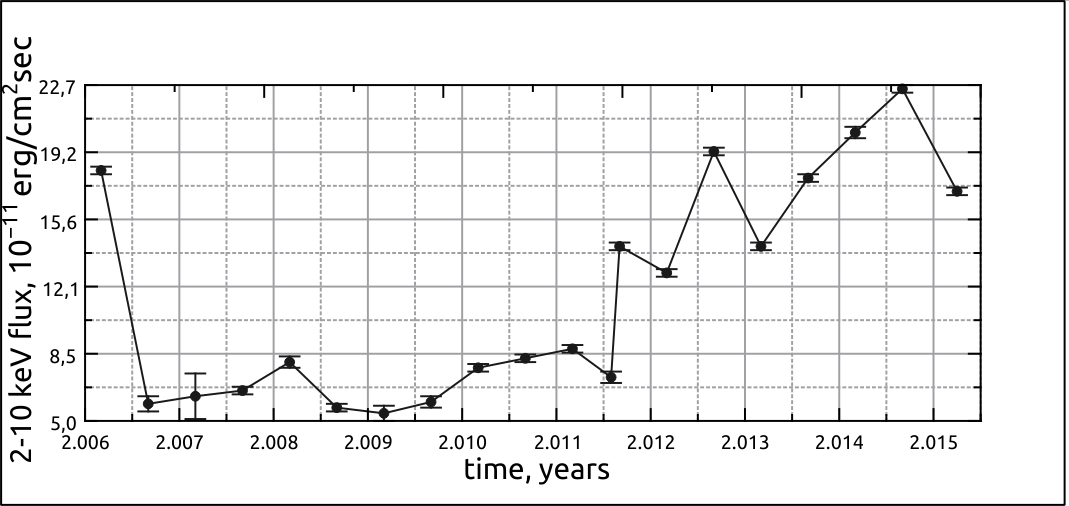}
      \caption{The 2-10 keV Suzaku/XIS light curve of NGC 1275.}
         \label{lcurve}
\end{figure*}

Based on this simple spectral model, using the \textit{flux} XSPEC command we calculated the AGN fluxes within the 2--10 keV energy range (see Table\ref{fits}, the last column) and plotted the light curve (see Fig.\ref{lcurve}).

\begin{figure*}[!htb]
   \centering
   \includegraphics[width=0.75\hsize]{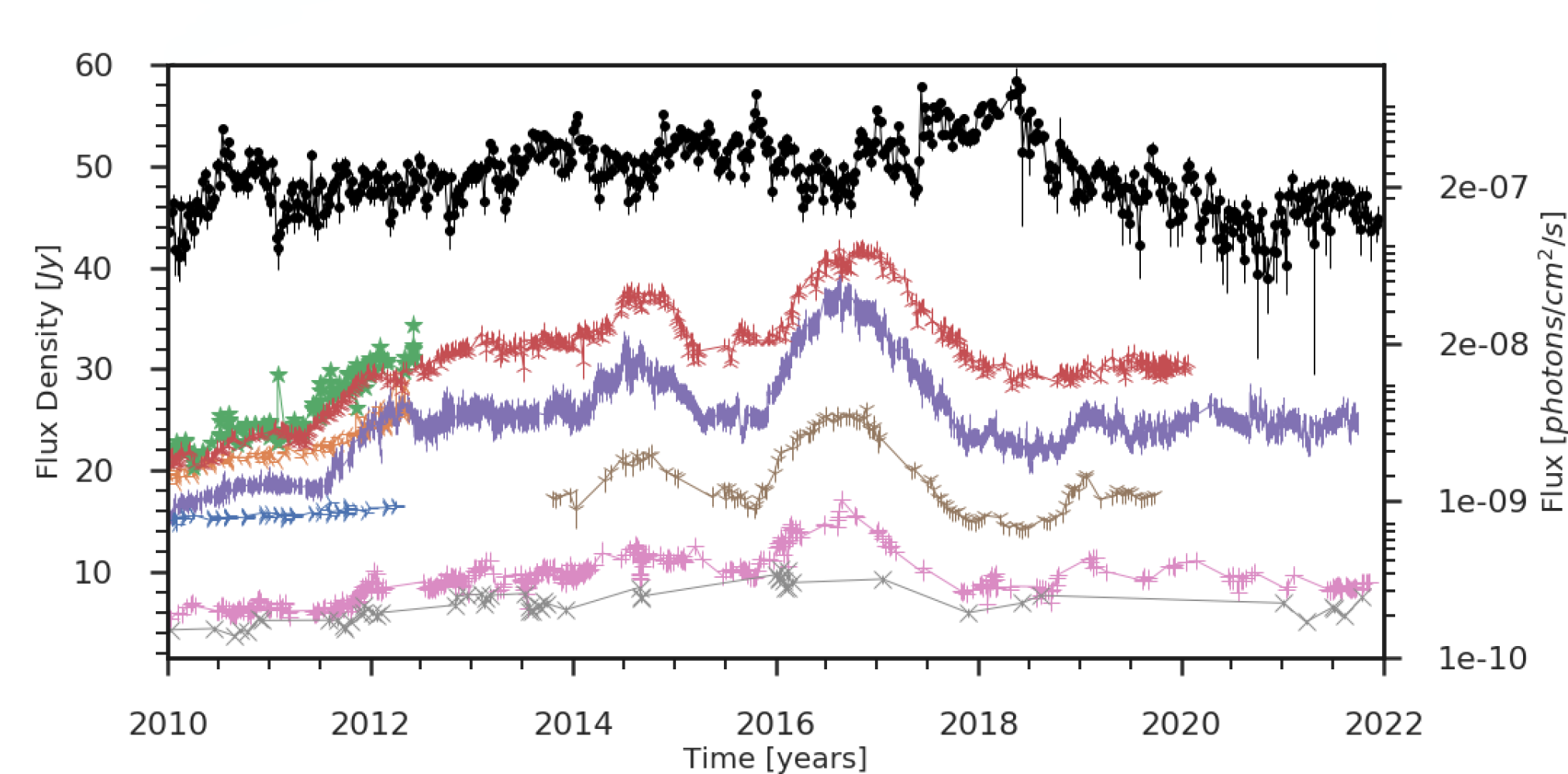}
   \includegraphics[width=.2\hsize]{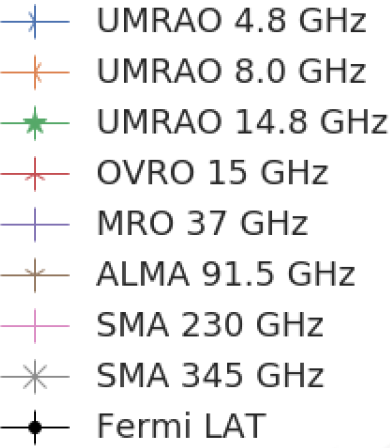}
      \caption{Variability light curves of NGC 1275. Radio light curves between 2010 and 2022 at number of frequencies 4.8, 8.0, 14.8 GHz were observed at the University of Michigan Radio Observatory; UMRAO, 15 GHz was observed at the Owen’s Valley Radio Observatory; OVRO, 37 GHz was observed at the Mets{\"a}hovi Radio Observatory; MRO), 91.5 GHz was observed at the Atacama Large Millimeter/submillimeter Array; ALMA, and 230, 345 GHz were observed at the Submillimeter Array; SMA. The black data set represents the $\gamma$-ray flux observed at the Fermi Large Area Telescope \citep{2009ApJ...699...31A}. Courtesy G. F. Paraschos et al. 2023 \citep{refId0}}
         \label{lcurve_2}
\end{figure*}

When examining the light curves in X-ray spectra in the period from 2010 to 2015, a consistent increase in flux density is observed. This matches the radio light curve demonstrating similar growth during the same period shown by Zhang et al. \citep{2022ApJ...925..207Z} and Paraschos et al. \citep{refId0} (see Fig.\ref{lcurve}, Fig.\ref{lcurve_2}). Notably, the X-ray emissions show good synchronization with emissions at lower frequencies, specifically at 15, 37, and 91.5 GHz. This suggests that radio and X-ray emissions likely originate from spatially proximate regions, potentially the base of the jet. However, this tie becomes less pronounced at higher frequencies, such as 230 and 345 GHz. There is no obvious correlation activity in X-ray and $\gamma$-ray ranges just considering from light curves. In the work by Paraschos et al. \citep{refId0} it was found that the $\gamma$-rays either precede the 230 GHz flux (345 GHz flux) by $\tau_{\gamma-230~GHz}=1.56\pm 0.27$ years ($\tau_{\gamma-345~GHz}= 1.57\pm0.49$ years) or trail the 230 GHz flux (345 GHz flux) by $\tau_{\gamma-230~GHz}=1.43\pm 0.30$ years ($\tau_{\gamma-345~GHz}= 1.58\pm0.64$ years). The authors' conclusion leans toward the $\gamma$-ray source being situated within the parsec-scale jet, positioned downstream of the core region of 3C 84. Also, Tanada et al.\citep{Tanada_2018} reported the flaring activity of NGC 1275 during the same period based on the Fermi observations. 

\section{Conclusions and further perspectives}
\label{sec4}

We have applied successfully our recipe to separate the cluster emission from active galactic nucleus emission in the uncleaned spectrum for all the available observational data by Suzaku/XIS within the energy 1-10 keV in a model-independent way. No spectral fitting models were applied to perform this spectral separation.  
As a result, we have the individual spectra for 20 periods of observations (the observations performed within a month were merged together). 

The spectral fitting of these spectra with the simplest power-law model with neutral absorption was performed. Despite this model is obviously not of high enough quality for these spectra (as it leads to the values of the discrepancy significantly higher than 1), it enables us to estimate the AGN 2-10 keV flux level at different periods of time and to disclose as the result that the 2-10 keV flux from the AGN nucleus in NGC 1275 were increasing significantly within the 2011-2015 period. 

Similar flux growth was detected within the same period in radio range \citep{2022ApJ...925..207Z} and at the highest $\gamma$-rays \citep{Tanada_2018}. This tells us that the observed flux growth could be rather induced by the activation of the processes in the jet than in the system "corona-accretion disk". So the next interesting point here would be to separate the jet and corona/disk spectral counterparts by applying for instance the method described in \citep{2020Univ....6..219F, 2022Galax..10....6F} to these Suzaku/XIS spectra together with the (quasi)simultaneous spectra on higher energies (i.e. Suzaku/HXD (Hard X-ray Detector)). This will give us the possibility to trace out the jet power in NGC 1275 and its evolution with time during the observational periods. The other intriguing possibility is to investigate the structure of the "central engine" of the NGC 1275 nucleus as there can be a double black hole \citep{2006MNRAS.366..758D,2019Galax...7...72B}.

The method itself and the scripts created for its realization can be used as well for another similar point-like object (AGN) surrounded by dense bright medium (cluster), under the condition that in the X-ray spectrum of the surroundings, there are present some bright features (such as emission or in some case absorption lines) not associated with the source of our interest (i.e. AGN). 


\section{Acknowledgement}
\label{sec6}

The team of authors is grateful to B.I. Hnatyk for the useful discussions and advice. L. Zadorozhna is grateful to Dr. Chunshan Lin for his hospitality at the Institute for Theoretical Physics, Jagiellonian University. L. Zadorozhna gratefully acknowledges ﬁnancial support from the National Science Center, Poland (Narodowe Centrum Nauki), grants No. UMO-2021/40/C/ST9/00015, UMO-2018/30/Q/ST9/00795. 
This research was made with the support of the Center for the Collective Use of Scientiﬁc Equipment ”Laboratory of High Energy Physics and Astrophysics” of Taras Shevchenko National University of Kyiv.


\section{Appendix A. The observation data LOG.}
\label{sec7}

See Table \ref{xislog}.

\begin{table}
\centering
    \caption{Suzaku/XIS observations LOG. The data events from various observations shown in one cell, more than one observation during one month were merged to produce one common spectrum from them marked as "total" below them.}
    \begin{tabular}{|p{50pt}|p{50pt}|p{80pt}|p{40pt}|p{80pt}|} \hline
        obsID & obs. & active & obs.time,& total flux \\
        & date & cameras & ksec & cts \\\hline
        800010010 & 2006-02-01 & XIS0-XIS3 & 50.4 & 417100$\pm$2131 \\\hline 
        101012010 & 2006-08-29 & XIS0-XIS3 & 150.9 & 13396000$\pm$3940 \\\hline
        101012020 & 2007-02-05 & XIS0, XIS1, XIS3 & 43.8 &  2884200$\pm$1778\\\hline
        102011010 & 2007-08-15 & XIS0, XIS1, XIS3 & 42.3 &  2889000$\pm$1777\\\hline
        102012010  & 2008-02-07 & XIS0, XIS1, XIS3 & 61.7 & 4560200$\pm$2235\\\hline
        103004010 & 2008-08-13 & XIS0, XIS1, XIS3 & 40.6 &  2431000$\pm$1635\\
        103005010 & 2008-08-14 & XIS0, XIS1, XIS3 & 21.5 &  517280$\pm$800\\
        Total & & & 61.6 & 2948280$\pm$1790 \\\hline
        103004020 & 2009-02-11 & XIS0, XIS1, XIS3 & 50.0 & 3265086$\pm$1886\\
        103005020 & 2009-02-12 & XIS0, XIS1, XIS3 & 28.8 & 1833000$\pm$1481\\
        Total & & & 78.8 & 5098086$\pm$2406\\\hline
        104018010 & 2009-08-26 & XIS0, XIS1, XIS3 & 41.3 &  2945544$\pm$1764\\
        104020010 & 2009-08-27 & XIS0, XIS1, XIS3 & 55.0 & 3945960$\pm$2127\\
        Total  & & & 96.3 &  6891504$\pm$2780\\\hline
        104019010 & 2010-02-01 & XIS0, XIS1, XIS3 & 38.6 & 2428740$\pm$1646\\
        104021010 & 2010-02-02 & XIS1, XIS3 & 21.6 & 923236$\pm$1045\\
        Total & & & 60.2 & 3351976$\pm$1922\\\hline
        105009010 & 2010-08-09 & XIS0, XIS1, XIS3 & 33.6 & 2500000$\pm$1697 \\
        105010010 & 2010-08-10 & XIS0, XIS1, XIS3 & 27.4 & 2125629$\pm$1576 \\
        Total & & & 61.0 & 4623629$\pm$2232\\\hline
        105010020 & 2011-02-02 & XIS0, XIS1, XIS3 & 21.1 & 1414400$\pm$1191 \\
        105009020 & 2011-02-03 & XIS0, XIS1, XIS3 & 40.5 & 2686704$\pm$1807 \\
        105028010 & 2011-02-21 & XIS0, XIS1, XIS3 & 20.6 & 1414872$\pm$1234 \\
        105027010 & 2011-02-22 & XIS0, XIS1, XIS3 & 45.3 & 2884020$\pm$1704 \\
        Total & & & 127.5 & 8399996$\pm$2968\\\hline
        106006010 & 2011-07-26 & XIS0, XIS1, XIS3 & 40.1 & 1767084$\pm$1452 \\
        106005010 & 2011-07-27 & XIS0, XIS1 & 40.8 & 1739514$\pm$1497 \\
        Total & & & 80.9 & 3506598$\pm$2045\\\hline
        106008010 & 2011-08-22 & XIS0, XIS1, XIS3 & 23.2 & 1607172$\pm$1224\\
        106007010 & 2011-08-23 & XIS0, XIS1, XIS3 & 21.0 & 1294568$\pm$1241\\
        Total & & & 44.2 & 2901740$\pm$1712 \\\hline
        106005020 & 2012-02-07 & XIS0, XIS1, XIS3& 46.8 & 3168472$\pm$1923 \\
        106007020 & 2012-02-08 & XIS0, XIS1, XIS3 & 20.9 & 1348127$\pm$1277\\
        Total & & & 76.7 & 4516599$\pm$2222\\\hline
        107006010 & 2012-08-20 & XIS0, XIS1, XIS3 & 23.7 & 643610$\pm$870\\
        107005010 & 2012-08-20 & XIS0, XIS1, XIS3 & 41.1 & 2855320$\pm$1808\\
        Total & & & 64.8 & 3496118$\pm$2006\\\hline
        107005020 & 2013-02-11 & XIS0, XIS1, XIS3 & 37.7 & 2715459$\pm$1749\\
        107006020 & 2013-02-12 & XIS0, XIS1, XIS3 & 22.0 & 594321$\pm$826\\
        Total & & & 59.7 & 3308823$\pm$1934\\\hline
        108005010 & 2013-08-15 & XIS0, XIS1, XIS3 & 41.3 & 2793686$\pm$1769 \\
        108006010 & 2013-08-16 & XIS0, XIS1, XIS3 & 21.6 & 1833876$\pm$1642 \\
        Total & & & 62.9 & 4627562$\pm$2369\\\hline
        108005020 & 2014-02-05 & XIS0, XIS1, XIS3 & 38.0 & 2544166$\pm$1691 \\
        108006020 & 2014-02-06 & XIS0, XIS1, XIS3 & 19.0 & 1302468$\pm$1213\\
        Total & & & 57.0 & 3846634$\pm$2017 \\\hline
        109005010 & 2014-08-27 & XIS0, XIS1, XIS3 & 20.0 & 1403352$\pm$1282 \\\hline
        109005020 & 2015-03-03 & XIS0, XIS1, XIS3 & 37.2 & 2412816$\pm$2167\\
        109006010 & 2015-03-04 & XIS0, XIS1, XIS3 & 23.6 & 1296420$\pm$1476\\
        109007010 & 2015-03-05 & XIS0, XIS1 & 30.7 & 3560711$\pm$2017\\
        Total & & & 91.5 & 7269947$\pm$3628\\\hline
    \end{tabular}
    \label{xislog}
\end{table}

\section{Appendix B. The emission lines.}
\label{sec8}

See Table \ref{apeclines0}, Table \ref{apeclines1}, Table \ref{apeclines2}, Table \ref{apeclines3}, Table \ref{apeclines4}. 

\begin{table}
\centering
    \caption{The emission lines in the collisionally-ionized diffuse gas spectrum following the atomicDB (APEC set) within the energy range 0.9-1.3 keV.}
    \begin{tabular}{|p{40pt}|p{55pt}|p{80pt}|p{60pt}|p{70pt}|} \hline
        ion & line energy, keV & emissivity, 10$^{-18}$ phot*cm$^3$s$^{-1}$ & T$_e$ peak, 10$^{+07}$ K & relative intensity  \\\hline
        Fe XXIV & 1.168 & 110.4 & 1.995 & 0.21 \\
        Fe XXIV & 1.163 & 57.33 & 1.995 & 0.11 \\
        Fe XIX & 1.146 & 50.02 & 1.000 & 0.10 \\
        Fe XXIII & 1.129 & 117.9 & 1.585 & 0.22 \\
        Fe XXIII & 1.125 & 75.68 & 1.585 & 0.14 \\
        Fe XXIV & 1.124 & 78.47 & 1.995 & 0.15 \\
        Fe XVII & 1.114 & 72.13 & 0.631 & 0.14 \\
        Fe XXIV & 1.109 & 140.8 & 1.995 & 0.27 \\
        Fe XVII & 1.102 & 100.1 & 0.631 & 0.19 \\
        Fe XVIII & 1.085 & 61.01 & 0.794 & 0.12 \\
        Fe XXIV & 1.085 & 70.43 & 1.995 & 0.13 \\
        Fe XVIII & 1.076 & 61.04 & 0.794 & 0.12 \\
        Ne IX & 1.074 & 50.44 & 0.398 & 0.10 \\
        Fe XXIII & 1.056 & 256.6 & 1.585 & 0.49 \\
        Fe XXII & 1.053 & 250.6 & 1.259 & 0.48 \\
        Fe XVII & 1.023 & 243.3 & 0.631 & 0.46 \\
        Ne X & 1.022 & 265.7 & 0.631 & 0.51 \\
        Ne X & 1.021 & 132.9 & 0.631 & 0.25 \\
        Fe XXIII & 1.020 & 138.7 & 1.585 & 0.26 \\
        Fe XVII & 1.011 & 222.4 & 0.631 & 0.42 \\
        Fe XXI & 1.009 & 524.3 & 1.259 & 1.00 \\
        Fe XXI & 1.000 & 91.68 & 1.259 & 0.17 \\
        Ni XIX & 0.997 & 72.07 & 0.794 & 0.14 \\
        Fe XXII & 0.972 & 88.00 & 1.259 & 0.17 \\
        Fe XXI & 0.967 & 64.42 & 1.259 & 0.12 \\
        Fe XX & 0.967 & 98.03 & 1.00 & 0.19 \\
        Fe XX & 0.965 & 228.0 & 1.00 & 0.43 \\
        Fe XX & 0.964 & 205.2 & 1.00 & 0.39 \\
        Fe XX & 0.960 & 56.13 & 1.00 & 0.11 \\
        Fe XX & 0.956 & 66.56 & 1.00 & 0.13 \\
        Fe XX & 0.949 & 54.71 & 1.00 & 0.10 \\
        Ne IX & 0.922 & 395.4 & 0.398 & 0.75 \\
        Fe XIX & 0.919 & 185.1 & 1.00 & 0.35 \\
        Fe XXI & 0.918 & 110.7 & 1.259 & 0.21 \\
        Fe XIX & 0.917 & 391.3 & 1.000 & 0.75 \\
        Ne IX & 0.915 & 72.31 & 0.398 & 0.14 \\
        Fe XIX & 0.909 & 62.26 & 1.00 & 0.12 \\
        Ne IX & 0.905 & 257.0 & 0.398 & 0.49 \\
        Fe XIX & 0.905 & 63.21 & 1.00 & 0.12 \\\hline
    \end{tabular}
    \vspace{0.05in} \label{apeclines0}
\end{table}

\begin{table}
\centering
    \caption{The emission lines in the collisionally-ionized diffuse gas spectrum following the atomicDB (APEC set) within the energy range 1.3-1.6 keV.}
    \begin{tabular}{|p{40pt}|p{55pt}|p{80pt}|p{60pt}|p{70pt}|} \hline
        ion & line energy, keV & emissivity, 10$^{-18}$ phot*cm$^3$s$^{-1}$ & T$_e$ peak, keV & relative intensity  \\\hline
        Mg XI & 1.579 & 13.63 & 0.5437 & 0.13 \\
        Fe XXIV & 1.553 & 15.61 & 1.719 & 0.15 \\
        Fe XXIII & 1.493 & 21.57 & 1.366 & 0.20 \\
        Fe XXIV & 1.491 & 15.45 & 1.719 & 0.14 \\
        Mg XII & 1.473 & 72.39 & 0.8617 & 0.68 \\
        Mg XII & 1.472 & 36.44 & 0.8617 & 0.34 \\
        Fe XXI & 1.446 & 17.11 & 1.085 & 0.16 \\
        Fe XXIII & 1.407 & 24.91 & 1.366 & 0.23 \\
        Fe XXII & 1.381 & 26.76 & 1.085 & 0.25 \\
        Mg XI & 1.352 & 106.8 & 0.5437 & 1.00 \\
        Mg XI & 1.343 & 18.67 & 0.5437 & 0.17\\
        Mg XI & 1.331 & 63.38 & 0.5437 & 0.59\\
        Fe XXI & 1.314 & 59.98 & 1.085 & 0.56\\\hline
    \end{tabular}
    \vspace{0.05in} \label{apeclines1}
\end{table}

\begin{table}
\centering
    \caption{The emission lines in the collisionally-ionized diffuse gas spectrum following the atomicDB (APEC set) within the energy range 1.8-2.2 keV.}
    \begin{tabular}{|p{40pt}|p{55pt}|p{80pt}|p{60pt}|p{70pt}|} \hline
        ion & line energy, keV & emissivity, 10$^{-18}$ phot*cm$^3$s$^{-1}$ & T$_e$ peak, keV & relative intensity  \\\hline
       Si XIII  &	2.183 & 11.41 &	0.8617 & 0.13 \\
       Si XIV 	& 2.006 & 52.34 & 1.366 & 0.60 \\
       Si XIV 	& 2.004 & 26.37 & 1.366 & 0.30 \\
       Si XIII & 1.865 & 86.66 & 0.8617 &	1.00 \\
       Si XIII & 1.854 & 14.31	& 86.17 & 0.17 \\
       Si XIII & 1.839 & 45.79 &	0.8617 & 0.53 \\\hline
    \end{tabular}
    \vspace{0.05in} \label{apeclines2}
\end{table}

\begin{table}
\centering
    \caption{The emission lines in the collisionally-ionized diffuse gas spectrum following the atomicDB (APEC set) within the energy range 2.4-2.6 keV.}
    \begin{tabular}{|p{40pt}|p{55pt}|p{80pt}|p{60pt}|p{70pt}|} \hline
        ion & line energy, keV & emissivity, 10$^{-18}$ phot*cm$^3$s$^{-1}$ & T$_e$ peak, keV & relative intensity  \\\hline
       S XVI & 2.623 & 19.50 & 2.165 & 0.58 \\
       S XVI & 2.620 & 9.770 & 2.165 & 0.29 \\
       S XV & 2.461 & 33.35 & 1.366 & 1.00 \\
       S XV & 2.447 & 5.136 & 1.085 & 0.15 \\
       S XV & 2.430 & 14.82 & 1.085 & 0.44 \\
       Si XIV & 2.377 & 7.051 & 1.366 & 0.21 \\
       Si XIV & 2.376 & 3.532 & 1.366 & 0.11 \\
       Si XIII & 2.294 & 3.744 & 0.8617 & 0.11\\\hline
    \end{tabular}
    \vspace{0.05in} \label{apeclines3}
\end{table}

\begin{table}
\centering
    \caption{The emission lines in the collisionally-ionized diffuse gas spectrum following the atomicDB (APEC set) within the energy range 6.5-7.0 keV.}
    \begin{tabular}{|p{40pt}|p{55pt}|p{80pt}|p{60pt}|p{70pt}|} \hline
        ion & line energy, keV & emissivity, 10$^{-18}$ phot*cm$^3$s$^{-1}$ & T$_e$ peak, keV & relative intensity  \\\hline
       Fe XXVI & 6.973 & 24.77 & 10.85 & 0.55 \\
       Fe XXVI & 6.952 & 12.51 & 10.85 & 0.28 \\
       Fe XXV & 6.700 & 45.26 & 5.437 & 1.00 \\
       Fe XXV & 6.682 & 8.213 & 5.437 & 0.18 \\
       Fe XXV & 6.668 & 9.056 & 5.437 & 0.20 \\
       Fe XXV & 6.637 & 14.23 & 5.437 & 0.31\\\hline
    \end{tabular}
    \vspace{0.05in} \label{apeclines4}
\end{table}

\url{http://www.atomdb.org/Webguide/webguide.php}

\section{Appendix C. Double background-subtracting.}
\label{sec9}

The algorithm we have used here to perform the double background-subtracting of the spectra of the AGN in NGC 1275 includes the following consequence of steps: 
\begin{itemize}
    \item{regular reduction of the observational data to collect the remote background, cluster, and AGN+cluster spectral counts;}
    \item{creating and rebinning of the remote background, cluster, and AGN+cluster spectral;}
    \item {extraction of the remote background from the cluster spectrum using our python script cluss\_corr.py shown in Appendix \ref{sec10}, applying the following console command to launch it:\\
    python3 clus\_corr.py -c clus\_raw.pha -b back.pha -c 1.0 clus\_clean.pha\\
    here clus\_raw.pha is an input uncleaned cluster spectrum, back.pha is a remote background spectrum, 1.0 in our case is the relation between the spaces of the cluster and background regions $S_{cluster}/S_{bkg}$, and clus\_clean.pha is an output file;}
    \item{use the resulting AGN+cluster spectrum with the cleaned cluster spectrum as a correction file XSPEC:\\
     \quad da source.grp\\
     \quad corfile clus\_clean.pha\\
     \quad cornorm 1.5\\}
     \item{set the "cornorm" coefficient to the value obtained from the relations between the amplitudes of the emission lines from AGN+cluster and cluster regions;}
     \item{add the systematic errors to the resulting spectra. As we have subtracted the remote background from the cluster spectrum, the additional errors are equivalent to the cluster + remote background errors. Depending on the observation, in our case, it's on the level of 0.07-0.2.}
\end{itemize} 

\section{Appendix C. back\_corr.py}
\label{sec10}

The contains of the back corrector:\\
\noindent from astropy.io import fits \\
\noindent import argparse as arg\\
\\
\noindent def parsing\_args():  \\                                                                           
\noindent \noindent   parser = arg.ArgumentParser()\\
\noindent \noindent   parser.add\_argument('-c', '--coeff', help="coefficient", dest="coeff", required = True)\\
\noindent \noindent  parser.add\_argument('-r', '--reno', help="spaces renormalization", dest="renorm")\\ 
\noindent \noindent   parser.add\_argument('-s', '--clus', help="cluster file", dest="clus\_file")\\
\noindent \noindent   parser.add\_argument('-b','--back', help="background file", dest="back\_file")\\
\noindent \noindent   parser.add\_argument('-o','--output',help="output file", dest="out\_file")\\
\noindent \noindent  args = parser.parse\_args()\\
\noindent    return args\\
\\
\noindent args = parsing\_args()\\
\noindent K = float(args.coeff)\\
\noindent R = float(args.renorm)\\

\#open the source (AGN+cluster), surroundings (cluster) and remote background FITS files\\
\\
\noindent hdub = fits.open(back\_file)\\
\noindent hduc = fits.open(clus\_file)\\
\\
\noindent bkg\_data = hdub[1].data\\
\noindent clus\_data = hduc[1].data\\

\#set the last (wrong) count to 0\\
\\
\noindent bkg\_data['COUNTS'][-1] = 0\\
\noindent clus\_data['COUNTS'][-1] = 0\\

\#set coefficients
\\
\noindent Kmax = len(clus\_data['COUNTS'])\\

\#rescaling the data and subtracting the surroundings spectrum from the source spectrum\\
\\
\noindent for i in range(0,Kmax,1):\\
\noindent \noindent    bkg\_data['COUNTS'][i] = clus\_data['COUNTS'][i]*K*R + bkg\_data['COUNTS'][i]*(1-K)\\

\#write the new FITS files with the rescaled source and remote background counts\\
\\
\noindent hdub.writeto(out\_file)\\



\footnotesize
\begin{thebibliography}{999}

\bibitem[{V{\'e}ron-Cetty} and {V{\'e}ron}(2010)]{Veron2010}
{V{\'e}ron-Cetty}, M.P.; {V{\'e}ron}, P.
\newblock {A catalogue of quasars and active nuclei: 13th edition}.
\newblock {\em Astron. Astrophys.} {\bf 2010}, {\em 518},~A10.
\newblock {\url{https://doi.org/10.1051/0004-6361/201014188}}.

\bibitem[{Buttiglione} et~al.(2010){Buttiglione}, {Capetti}, {Celotti}, {Axon},
  {Chiaberge}, {Macchetto}, and {Sparks}]{Buttiglione2010}
{Buttiglione}, S.; {Capetti}, A.; {Celotti}, A.; {Axon}, D.J.; {Chiaberge}, M.;
  {Macchetto}, F.D.; {Sparks}, W.B.
\newblock {An optical spectroscopic survey of the 3CR sample of radio galaxies
  with z < 0.3. II. Spectroscopic classes and accretion modes in radio-loud
  AGN}.
\newblock {\em Astron. Astrophys.} {\bf 2010}, {\em 509},~A6,
  \href{http://xxx.lanl.gov/abs/0911.0536}{{\normalfont
  [arXiv:astro-ph.CO/0911.0536]}}.
\newblock {\url{https://doi.org/10.1051/0004-6361/200913290}}.

\bibitem[{Vermeulen} et~al.(1994){Vermeulen}, {Readhead}, and
  {Backer}]{Vermeulen1994}
{Vermeulen}, R.C.; {Readhead}, A.C.S.; {Backer}, D.C.
\newblock {Discovery of a Nuclear Counterjet in NGC 1275: A New Way to Probe
  the Parsec-Scale Environment}.
\newblock {\em Astrophys. J.} {\bf 1994}, {\em 430},~L41.
\newblock {\url{https://doi.org/10.1086/187433}}.

\bibitem[{Falco} and et~al(1999)]{1999PASP..111..438F}
{Falco}, E.E.; et~al.
\newblock {The Updated Zwicky Catalog (UZC)}.
\newblock {\em PASP} {\bf 1999}, {\em 111},~438--452,
  \href{http://xxx.lanl.gov/abs/astro-ph/9904265}{{\normalfont
  [arXiv:astro-ph/astro-ph/9904265]}}.
\newblock {\url{https://doi.org/10.1086/316343}}.

\bibitem[Wilman et~al.(2005)Wilman, Edge, and Johnstone]{Wilman2005}
Wilman, R.J.; Edge, A.C.; Johnstone, R.M.
\newblock {The nature of the molecular gas system in the core of NGC 1275}.
\newblock {\em MNRAS} {\bf 2005}, {\em 359},~755--764,
  \href{http://xxx.lanl.gov/abs/https://academic.oup.com/mnras/article-pdf/359/2/755/3108907/359-2-755.pdf}{{\normalfont
  [https://academic.oup.com/mnras/article-pdf/359/2/755/3108907/359-2-755.pdf]}}.
\newblock {\url{https://doi.org/10.1111/j.1365-2966.2005.08956.x}}.

\bibitem[{Lynds}(1970)]{1970ApJ...159L.151L}
{Lynds}, R.
\newblock {Improved Photographs of the NGC 1275 Phenomenon}.
\newblock {\em Astrophys. J.} {\bf 1970}, {\em 159},~L151--L154.
\newblock {\url{https://doi.org/10.1086/180500}}.

\bibitem[{Cobos} et~al.(2018){Cobos}, {Rich}, and {Great Observatories All-sky
  LIRG Survey (GOALS)}]{2018AAS...23125211C}
{Cobos}, A.S.; {Rich}, J.; {Great Observatories All-sky LIRG Survey (GOALS)}.
\newblock {Mapping the filaments in NGC 1275}.
\newblock In Proceedings of the American Astronomical Society Meeting Abstracts
  \#231,  2018, Vol. 231, {\em American Astronomical Society Meeting
  Abstracts}, p. 252.11.

\bibitem[{Fabian} and et~al(1974)]{1974ApJ...189L..59F}
{Fabian}, A.C.; et~al.
\newblock {Copernicus X-Ray Observations of NGC 1275 and the Core of the
  Perseus Cluster}.
\newblock {\em Astrophys. J.} {\bf 1974}, {\em 189},~L59.
\newblock {\url{https://doi.org/10.1086/181464}}.

\bibitem[{Fabian} and et~al(2011)]{2011MNRAS.418.2154F}
{Fabian}, A.C.; et~al.
\newblock {A wide Chandra view of the core of the Perseus cluster}.
\newblock {\em MNRAS} {\bf 2011}, {\em 418},~2154--2164,
  \href{http://xxx.lanl.gov/abs/1105.5025}{{\normalfont
  [arXiv:astro-ph.CO/1105.5025]}}.
\newblock {\url{https://doi.org/10.1111/j.1365-2966.2011.19402.x}}.

\bibitem[{Salom{\'e}} and at~al(2008)]{2008A&A...484..317S}
{Salom{\'e}}, P.; at~al.
\newblock {Cold gas in the Perseus cluster core: excitation of molecular gas in
  filaments}.
\newblock {\em Astron. Astrophys.} {\bf 2008}, {\em 484},~317--325,
  \href{http://xxx.lanl.gov/abs/0804.2113}{{\normalfont
  [arXiv:astro-ph/0804.2113]}}.
\newblock {\url{https://doi.org/10.1051/0004-6361:200809493}}.

\bibitem[{Lim} et~al.(2008){Lim}, {Ao}, and
  {Dinh-V-Trung}]{2008ApJ...672..252L}
{Lim}, J.; {Ao}, Y.; {Dinh-V-Trung}.
\newblock {Radially Inflowing Molecular Gas in NGC 1275 Deposited by an X-Ray
  Cooling Flow in the Perseus Cluster}.
\newblock {\em Astrophys. J.} {\bf 2008}, {\em 672},~252--265,
  \href{http://xxx.lanl.gov/abs/0712.2979}{{\normalfont
  [arXiv:astro-ph/0712.2979]}}.
\newblock {\url{https://doi.org/10.1086/523664}}.

\bibitem[{Boehringer} et~al.(1993){Boehringer}, {Voges}, {Fabian}, {Edge}, and
  {Neumann}]{1993MNRAS.264L..25B}
{Boehringer}, H.; {Voges}, W.; {Fabian}, A.C.; {Edge}, A.C.; {Neumann}, D.M.
\newblock {A ROSAT HRI study of the interaction of the X-ray emitting gas and
  radio lobes of NGC 1275.}
\newblock {\em MNRAS} {\bf 1993}, {\em 264},~L25--L28.
\newblock {\url{https://doi.org/10.1093/mnras/264.1.L25}}.

\bibitem[{Sinitsyna} et~al.(2013){Sinitsyna}, {Nikolsky}, and {Y
  Sinitsyna}]{2013JPhCS.409a2111S}
{Sinitsyna}, V.G.; {Nikolsky}, S.I.; {Y Sinitsyna}, V.
\newblock {Long-term observation of Seyfert Galaxies NGC 1275 and 3C 382 at TeV
  energies by SHALON}.
\newblock In Proceedings of the Journal of Physics Conference Series,  2013,
  Vol. 409, {\em Journal of Physics Conference Series}, p. 012111.
\newblock {\url{https://doi.org/10.1088/1742-6596/409/1/012111}}.

\bibitem[200(2003)]{2003cxo..pres...17.}
{Chandra ``Hears'' A Black Hole For The First Time}.
\newblock Chandra Press Release,  2003.

\bibitem[{Fabian} and et~al(2003)]{2003MNRAS.344L..43F}
{Fabian}, A.C.; et~al.
\newblock {A deep Chandra observation of the Perseus cluster: shocks and
  ripples}.
\newblock {\em MNRAS} {\bf 2003}, {\em 344},~L43--L47,
  \href{http://xxx.lanl.gov/abs/astro-ph/0306036}{{\normalfont
  [arXiv:astro-ph/astro-ph/0306036]}}.
\newblock {\url{https://doi.org/10.1046/j.1365-8711.2003.06902.x}}.

\bibitem[{Baghmanyan} et~al.(2017){Baghmanyan}, {Gasparyan}, and
  {Sahakyan}]{2017ApJ...848..111B}
{Baghmanyan}, V.; {Gasparyan}, S.; {Sahakyan}, N.
\newblock {Rapid Gamma-Ray Variability of NGC 1275}.
\newblock {\em Astrophys. J.} {\bf 2017}, {\em 848},~111,
  \href{http://xxx.lanl.gov/abs/1709.03755}{{\normalfont
  [arXiv:astro-ph.HE/1709.03755]}}.
\newblock {\url{https://doi.org/10.3847/1538-4357/aa8c7b}}.

\bibitem[{Kataoka} et~al.(2010){Kataoka}, {Stawarz}, {Cheung}, {Tosti},
  {Cavazzuti}, {Celotti}, {Nishino}, {Fukazawa}, {Thompson}, and
  {McConville}]{2010ApJ...715..554K}
{Kataoka}, J.; {Stawarz}, {\L}.; {Cheung}, C.C.; {Tosti}, G.; {Cavazzuti}, E.;
  {Celotti}, A.; {Nishino}, S.; {Fukazawa}, Y.; {Thompson}, D.J.; {McConville},
  W.F.
\newblock {$\gamma$-ray Spectral Evolution of NGC 1275 Observed with Fermi
  Large Area Telescope}.
\newblock {\em Astrophys. J.} {\bf 2010}, {\em 715},~554--560,
  \href{http://xxx.lanl.gov/abs/1004.2352}{{\normalfont
  [arXiv:astro-ph.HE/1004.2352]}}.
\newblock {\url{https://doi.org/10.1088/0004-637X/715/1/554}}.

\bibitem[Chitnis et~al.(2020)Chitnis, Shukla, Singh, Roy, Bhattacharyya,
  Chandra, and Stewart]{galaxies8030063}
Chitnis, V.; Shukla, A.; Singh, K.P.; Roy, J.; Bhattacharyya, S.; Chandra, S.;
  Stewart, G.
\newblock X-ray and Gamma-ray Variability of NGC 1275.
\newblock {\em Galaxies} {\bf 2020}, {\em 8}.
\newblock {\url{https://doi.org/10.3390/galaxies8030063}}.

\bibitem[{Imazato} et~al.(2021){Imazato}, {Fukazawa}, {Sasada}, and
  {Sakamoto}]{2021ApJ...906...30I}
{Imazato}, F.; {Fukazawa}, Y.; {Sasada}, M.; {Sakamoto}, T.
\newblock {Origin of the UV to X-Ray Emission of Radio Galaxy NGC 1275 Explored
  by Analyzing Its Variability}.
\newblock {\em Astrophys. J.} {\bf 2021}, {\em 906},~30,
  \href{http://xxx.lanl.gov/abs/2011.10299}{{\normalfont
  [arXiv:astro-ph.HE/2011.10299]}}.
\newblock {\url{https://doi.org/10.3847/1538-4357/abc7bc}}.

\bibitem[{Aleksi{\'c}} and et~al(2014)]{2014A&A...564A...5A}
{Aleksi{\'c}}, J.; et~al.
\newblock {Contemporaneous observations of the radio galaxy NGC 1275 from radio
  to very high energy $\gamma$-rays}.
\newblock {\em Astron. Astrophys.} {\bf 2014}, {\em 564},~A5,
  \href{http://xxx.lanl.gov/abs/1310.8500}{{\normalfont
  [arXiv:astro-ph.HE/1310.8500]}}.
\newblock {\url{https://doi.org/10.1051/0004-6361/201322951}}.

\bibitem[{Nesterov} et~al.(1995){Nesterov}, {Lyuty}, and
  {Valtaoja}]{1995A&A...296..628N}
{Nesterov}, N.S.; {Lyuty}, V.M.; {Valtaoja}, E.
\newblock {Radio and optical evolution of the Seyfert galaxy NGC 1275.}
\newblock {\em Astron. Astrophys.} {\bf 1995}, {\em 296},~628.

\bibitem[{Fabian} et~al.(2015){Fabian}, {Walker}, {Pinto}, {Russell}, and
  {Edge}]{2015MNRAS.451.3061F}
{Fabian}, A.C.; {Walker}, S.A.; {Pinto}, C.; {Russell}, H.R.; {Edge}, A.C.
\newblock {Effects of the variability of the nucleus of NGC 1275 on X-ray
  observations of the surrounding intracluster medium}.
\newblock {\em MNRAS} {\bf 2015}, {\em 451},~3061--3067,
  \href{http://xxx.lanl.gov/abs/1505.03754}{{\normalfont
  [arXiv:astro-ph.HE/1505.03754]}}.
\newblock {\url{https://doi.org/10.1093/mnras/stv1134}}.

\bibitem[{Fukazawa} and et~al(2018)]{2018ApJ...855...93F}
{Fukazawa}, Y.; et~al.
\newblock {X-Ray and GeV Gamma-Ray Variability of the Radio Galaxy NGC 1275}.
\newblock {\em Astrophys. J.} {\bf 2018}, {\em 855},~93.
\newblock {\url{https://doi.org/10.3847/1538-4357/aaabc0}}.

\bibitem[{Ahnen} and et~al(2016)]{2016A&A...589A..33A}
{Ahnen}, M.L.; et~al.
\newblock {Deep observation of the NGC 1275 region with MAGIC: search of
  diffuse $\gamma$-ray emission from cosmic rays in the Perseus cluster}.
\newblock {\em Astron. Astrophys.} {\bf 2016}, {\em 589},~A33,
  \href{http://xxx.lanl.gov/abs/1602.03099}{{\normalfont
  [arXiv:astro-ph.HE/1602.03099]}}.
\newblock {\url{https://doi.org/10.1051/0004-6361/201527846}}.

\bibitem[{Ansoldi} and et~al(2018)]{2018A&A...617A..91M}
{Ansoldi}, S.; et~al.
\newblock {Gamma-ray flaring activity of NGC1275 in 2016-2017 measured by
  MAGIC}.
\newblock {\em Astron. Astrophys.} {\bf 2018}, {\em 617},~A91,
  \href{http://xxx.lanl.gov/abs/1806.01559}{{\normalfont
  [arXiv:astro-ph.HE/1806.01559]}}.
\newblock {\url{https://doi.org/10.1051/0004-6361/201832895}}.

\bibitem[{Britzen} et~al.(2019){Britzen}, {Fendt}, {Zaja{\v{c}}ek}, {Jaron},
  {Pashchenko}, {Aller}, and {Aller}]{2019Galax...7...72B}
{Britzen}, S.; {Fendt}, C.; {Zaja{\v{c}}ek}, M.; {Jaron}, F.; {Pashchenko}, I.;
  {Aller}, M.F.; {Aller}, H.D.
\newblock {3C 84: Observational Evidence for Precession and a Possible Relation
  to TeV Emission}.
\newblock {\em Galaxies} {\bf 2019}, {\em 7},~72.
\newblock {\url{https://doi.org/10.3390/galaxies7030072}}.

\bibitem[{Aharonian} and et~al(2018)]{2018PASJ...70...13H}
{Aharonian}, F.; et~al.
\newblock {Hitomi observation of radio galaxy NGC 1275: The first X-ray
  microcalorimeter spectroscopy of Fe-K$\alpha$ line emission from an active
  galactic nucleus}.
\newblock {\em PASJ} {\bf 2018}, {\em 70},~13,
  \href{http://xxx.lanl.gov/abs/1711.06289}{{\normalfont
  [arXiv:astro-ph.HE/1711.06289]}}.
\newblock {\url{https://doi.org/10.1093/pasj/psx147}}.

\bibitem[{Yamazaki} and et~al(2013)]{2013PASJ...65...30Y}
{Yamazaki}, S.; et~al.
\newblock {X-Ray and Optical Monitoring of a Gamma-Ray-Emitting Radio Galaxy,
  NGC 1275}.
\newblock {\em PASJ} {\bf 2013}, {\em 65},~30.
\newblock {\url{https://doi.org/10.1093/pasj/65.2.30}}.

\bibitem[{Churazov} et~al.(2003){Churazov}, {Forman}, {Jones}, and
  {B{\"o}hringer}]{2003ApJ...590..225C}
{Churazov}, E.; {Forman}, W.; {Jones}, C.; {B{\"o}hringer}, H.
\newblock {XMM-Newton Observations of the Perseus Cluster. I. The Temperature
  and Surface Brightness Structure}.
\newblock {\em Astrophys. J.} {\bf 2003}, {\em 590},~225--237,
  \href{http://xxx.lanl.gov/abs/astro-ph/0301482}{{\normalfont
  [arXiv:astro-ph/astro-ph/0301482]}}.
\newblock {\url{https://doi.org/10.1086/374923}}.

\bibitem[{Nagai} and et~al(2019)]{Nagai2019}
{Nagai}, H.; et~al.
\newblock {The ALMA Discovery of the Rotating Disk and Fast Outflow of Cold
  Molecular Gas in NGC 1275}.
\newblock {\em Astrophys. J.} {\bf 2019}, {\em 883},~193,
  \href{http://xxx.lanl.gov/abs/1905.06017}{{\normalfont
  [arXiv:astro-ph.GA/1905.06017]}}.
\newblock {\url{https://doi.org/10.3847/1538-4357/ab3e6e}}.

\bibitem[{Reynolds} and at~al(2021)]{2021arXiv210804276R}
{Reynolds}, C.S.; at~al.
\newblock {Probing the circumnuclear environment of NGC1275 with
  High-Resolution X-ray spectroscopy}.
\newblock {\em arXiv e-prints} {\bf 2021}, p. arXiv:2108.04276,
  \href{http://xxx.lanl.gov/abs/2108.04276}{{\normalfont
  [arXiv:astro-ph.HE/2108.04276]}}.

\bibitem[{Gulati} et~al.(2021){Gulati}, {Bhattacharya}, {Bhattacharyya},
  {Bhatt}, {Stalin}, and {Agrawal}]{2021MNRAS.503..446G}
{Gulati}, S.; {Bhattacharya}, D.; {Bhattacharyya}, S.; {Bhatt}, N.; {Stalin},
  C.S.; {Agrawal}, V.K.
\newblock {Multiwavelength monitoring of NGC 1275 over a decade: evidence of a
  shift in synchrotron peak frequency and long-term multiband flux increase}.
\newblock {\em MNRAS} {\bf 2021}, {\em 503},~446--457,
  \href{http://xxx.lanl.gov/abs/2101.11540}{{\normalfont
  [arXiv:astro-ph.HE/2101.11540]}}.
\newblock {\url{https://doi.org/10.1093/mnras/stab244}}.

\bibitem[{Zhang} et~al.(2022){Zhang}, {Wang}, {Gurwell}, and
  {Wiita}]{2022ApJ...925..207Z}
{Zhang}, P.; {Wang}, Z.; {Gurwell}, M.; {Wiita}, P.J.
\newblock {A Double-period Oscillation Signal in Millimeter Emission of the
  Radio Galaxy NGC 1275}.
\newblock {\em Astrophys. J.} {\bf 2022}, {\em 925},~207,
  \href{http://xxx.lanl.gov/abs/2112.03544}{{\normalfont
  [arXiv:astro-ph.HE/2112.03544]}}.
\newblock {\url{https://doi.org/10.3847/1538-4357/ac425c}}.

\bibitem[{Joye} and {Mandel}(2003)]{2003ASPC..295..489J}
{Joye}, W.A.; {Mandel}, E.
\newblock {New Features of SAOImage DS9}.
\newblock In Proceedings of the Astronomical Data Analysis Software and Systems
  XII; {Payne}, H.E.; {Jedrzejewski}, R.I.; {Hook}, R.N., Eds.,  2003, Vol.
  295, {\em Astronomical Society of the Pacific Conference Series}, p. 489.

\bibitem[{Kettula} et~al.(2013){Kettula}, {Nevalainen}, and
  {Miller}]{2013A&A...552A..47K}
{Kettula}, K.; {Nevalainen}, J.; {Miller}, E.D.
\newblock {Cross-calibration of Suzaku/XIS and XMM-Newton/EPIC using galaxy
  clusters}.
\newblock {\em Astrophys. J.} {\bf 2013}, {\em 552},~A47,
  \href{http://xxx.lanl.gov/abs/1301.2947}{{\normalfont
  [arXiv:astro-ph.IM/1301.2947]}}.
\newblock {\url{https://doi.org/10.1051/0004-6361/201220408}}.

\bibitem[{Liedahl} et~al.(1995){Liedahl}, {Osterheld}, and
  {Goldstein}]{1995ApJ...438L.115L}
{Liedahl}, D.A.; {Osterheld}, A.L.; {Goldstein}, W.H.
\newblock {New Calculations of Fe L-Shell X-Ray Spectra in High-Temperature
  Plasmas}.
\newblock {\em Astrophys. J.} {\bf 1995}, {\em 438},~L115.
\newblock {\url{https://doi.org/10.1086/187729}}.

\bibitem[{Paraschos, G. F.} et~al.(2023){Paraschos, G. F.}, {Mpisketzis, V.},
  {Kim, J.-Y.}, {Witzel, G.}, {Krichbaum, T. P.}, {Zensus, J. A.}, {Gurwell, M.
  A.}, {L\"ahteenm\"aki, A.}, {Tornikoski, M.}, {Kiehlmann, S.}, and {Readhead,
  A. C. S.}]{refId0}
{Paraschos, G. F.}.; {Mpisketzis, V.}.; {Kim, J.-Y.}.; {Witzel, G.}.;
  {Krichbaum, T. P.}.; {Zensus, J. A.}.; {Gurwell, M. A.}.; {L\"ahteenm\"aki,
  A.}.; {Tornikoski, M.}.; {Kiehlmann, S.}.;  et~al.
\newblock A multi-band study and exploration of the radio wave- connection in
  3C 84.
\newblock {\em Astron. Astrophys.} {\bf 2023}, {\em 669},~A32.
\newblock {\url{https://doi.org/10.1051/0004-6361/202244814}}.

\bibitem[Tanada et~al.(2018)Tanada, Kataoka, Arimoto, Akita, Cheung, Digel, and
  Fukazawa]{Tanada_2018}
Tanada, K.; Kataoka, J.; Arimoto, M.; Akita, M.; Cheung, C.C.; Digel, S.W.;
  Fukazawa, Y.
\newblock The Origins of the Gamma-Ray Flux Variations of NGC 1275 Based on
  Eight Years of Fermi-LAT Observations.
\newblock {\em The Astrophysical Journal} {\bf 2018}, {\em 860},~74.
\newblock {\url{https://doi.org/10.3847/1538-4357/aac26b}}.

\bibitem[{Abdo} and et~al(2009)]{2009ApJ...699...31A}
{Abdo}, A.A.; et~al.
\newblock {Fermi Discovery of Gamma-ray Emission from NGC 1275}.
\newblock {\em Astrophys. J.} {\bf 2009}, {\em 699},~31--39,
  \href{http://xxx.lanl.gov/abs/0904.1904}{{\normalfont
  [arXiv:astro-ph.HE/0904.1904]}}.
\newblock {\url{https://doi.org/10.1088/0004-637X/699/1/31}}.

\bibitem[{Fedorova} et~al.(2020){Fedorova}, {Hnatyk}, {Zhdanov}, and {Del
  Popolo}]{2020Univ....6..219F}
{Fedorova}, E.; {Hnatyk}, B.I.; {Zhdanov}, V.I.; {Del Popolo}, A.
\newblock {X-ray Properties of 3C 111: Separation of Primary Nuclear Emission
  and Jet Continuum}.
\newblock {\em Universe} {\bf 2020}, {\em 6},~219.

\bibitem[{Fedorova} et~al.(2022){Fedorova}, {Hnatyk}, {Del Popolo},
  {Vasylenko}, and {Voitsekhovskyi}]{2022Galax..10....6F}
{Fedorova}, E.; {Hnatyk}, B.; {Del Popolo}, A.; {Vasylenko}, A.;
  {Voitsekhovskyi}, V.
\newblock {Non-Thermal Emission from Radio-Loud AGN Jets: Radio vs. X-rays}.
\newblock {\em Galaxies} {\bf 2022}, {\em 10},~6.
\newblock {\url{https://doi.org/10.3390/galaxies10010006}}.

\bibitem[{Dunn} et~al.(2006){Dunn}, {Fabian}, and
  {Sanders}]{2006MNRAS.366..758D}
{Dunn}, R.J.H.; {Fabian}, A.C.; {Sanders}, J.S.
\newblock {Precession of the super-massive black hole in NGC 1275 (3C 84)?}
\newblock {\em MNRAS} {\bf 2006}, {\em 366},~758--766,
  \href{http://xxx.lanl.gov/abs/astro-ph/0512022}{{\normalfont
  [arXiv:astro-ph/astro-ph/0512022]}}.
\newblock {\url{https://doi.org/10.1111/j.1365-2966.2005.09928.x}}.

\end{thebibliography}
\end{document}